%% file: Atomic_transitions_by_GW.tex
\documentclass[nofootinbib,eqsecnum,superscriptaddress,aps,11pt]{revtex4-2}

\usepackage{amsmath,amssymb,amsfonts}

\usepackage{bm} 
\usepackage{bbm} 
\usepackage{mathrsfs} 
\usepackage[makeroom]{cancel} 

\usepackage{footnotebackref} 

\newcommand{\abs}[1]{{\left|{#1}\right|}} 
\newcommand{\inner}[2]{{\langle {#1}\vert {#2} \rangle}} 
\newcommand{\ket}[1]{\vert{#1}\rangle} 
\newcommand{\bra}[1]{\langle{#1}\vert} 

\DeclareMathAlphabet{\mathpzc}{OT1}{pzc}{m}{it} 

\newcommand{\secref}[1]{Sec.~\ref{#1}}

\begin{document}
\title
{Atomic electron transitions of hydrogen-like atoms induced by gravitational waves}


\author{Bo-Hung Chen}
\email{kenny81778189@gmail.com}
\affiliation{Department of Physics, National Taiwan University, Taipei 10617, Taiwan}

\author{Dah-Wei Chiou}
\email{dwchiou@gmail.com}
\affiliation{Graduate Institute of Electronics Engineering, National Taiwan University, Taipei 10617, Taiwan}

\begin{abstract}
As a realistic model of a quantum system of matter, this paper investigates the gravitational-wave effects on a hydrogen-like atom. By formulating the tetrad formalism of linearized gravity, we naturally incorporate the gravitational-wave effects through minimal coupling in the covariant Dirac equation. The atomic electron transition rates induced by the gravitational wave are calculated using the first-order perturbation theory, revealing a distinctive selection rule along with Fermi's golden rule. This rule can be elegantly understood in terms of gravitons as massless spin-2 particles. Our results suggest the existence of gravitons and may lead to a novel approach to probe Ultra-High-Frequency Gravitational Waves (UHF-GWs).
\end{abstract}

\maketitle

\section{Introduction}
Understanding the effects of gravitational waves on matter presents a significant challenge and has caused considerable confusion. The difficulty arises partly because a well-defined localized energy-momentum for the gravitational field is impossible, obscuring the notion of energy carried by a gravitational wave (see Sections 20.4, 35.7, and 35.15 of \cite{misner1973gravitation}).
Conflicting opinions on whether gravitational waves exert measurable effects on matter or not persisted until Feynman \cite{Feynman:a,Feynman:b} and Bondi \cite{Bondi:1957dt} proposed the \emph{sticky bead argument}, finally concluding the issue in the affirmative.
The rationale of the sticky bead argument also applies to different classical mechanical systems, which can be leveraged as gravitational wave detectors (see Chapter 37 of \cite{misner1973gravitation}). A prominent example is the \emph{resonant mass detector}, which is employed in various operating experiments (see \cite{aguiar2011past} for a review) as an alternative to interferometric gravitational wave detectors (see \cite{saulson2018interferometric, reitze2019advanced} for reviews).

The response of a classical mechanical system to a gravitational wave can be understood in terms of the tidal force produced by the gravitational wave, which acts as the driving force against the interacting force (e.g., friction in the sticky bead system, elastic and damping forces in the resonant mass detector, etc.)\ between mass elements of the system (see Chapter 37, especially Section 37.2, of \cite{misner1973gravitation} for a detailed account).
This classical picture however may not reveal certain subtle effects that do not directly arise from the tidal force. Hence, for a more comprehensive understanding, we need to study the response of a quantum system to gravitational waves in a more fundamental context, rather than relying solely on the phenomenological approach for classical mechanical systems.
Serving as the simplest theoretical model of a quantum system coupled with gravity, the response of an Unruh--DeWitt detector to a gravitational wave background was investigated in \cite{PhysRevD.105.024053}, which unveils a novel effect beyond the scope of gravitational-wave tidal force.
More works have also been devoted to explore the gravitational-wave effects on the Unruh--DeWitt detector \cite{Prokopec_2023, barman2023entanglement}.
However, the reliability of the new effect is questionable, as the Unruh--DeWitt detector is a rather artificial model.

In this paper, we undertake a more realistic model --- a hydrogen-like atom (i.e.\ a hydrogen atom, a single-electron ion, or any atomic entity with a single electron seeing the rest of the entity effectively as a point charge such as a Rydberg atom) interacting with a gravitational wave.
We reformulate the theory of linearized gravity in the tetrad formalism. This enables us to naturally incorporate gravitational-wave effects through minimal coupling in the covariant Dirac equation for the electron in a hydrogen-like atom. Employing the first-order perturbation theory, we then compute the atomic electron transition rates induced by the gravitational wave.
The transition rates reveal an intriguing selection rule along with Fermi's golden rule, which can be elegantly explained in terms of gravitons as massless spin-2 particles.
Our results mark a considerable stride in our understanding of gravitational-wave interactions with atomic systems and strongly suggest the existence of gravitons, even though the gravitational wave is treated purely as a classical field in our approach.
As our analysis is grounded in fundamental assumptions and does not rely on any quantization of gravitational fields, our results remain viable regardless of the existence of gravitons.

The main objective of this work is to understand theoretical aspects of gravitational-wave effects on matter. In experimental aspects, measuring the atomic electron transitions induced by gravitational waves is virtually impossible on Earth, as a gravitational wave is extremely weak upon arrival. In outer space, however, gravitational waves can be sufficiently strong, leaving detectable traces on the interstellar medium. By carefully analyzing the emission or absorption spectra of hydrogen-like atoms in the interstellar medium, our research may lead to a novel approach to probe Ultra-High-Frequency Gravitational Waves (UHF-GWs) \cite{aggarwal2021challenges}. UHF-GWs are anticipated to unveil new physics beyond the Standard Model, but only a handful of detector concepts have been proposed so far for their measurement \cite{aggarwal2021challenges,domcke2023electromagnetic}.

This paper is organized as follows.
In \secref{sec:tetrad formalism}, we cast the linearized theory of gravity in the tetrad formalism.
In \secref{sec:hydrogen-like atoms}, we formulate the covariant Dirac equation for the electron in a hydrogen-like atom subject to a gravitational wave.
In \secref{sec:atomic transitions}, employing the first-order perturbation theory, we then compute the atomic electron transition rates induced by a gravitational wave.
In \secref{sec:summary}, the main results are summarized and their implications are discussed.

The notation and convention are adopted as follows.
The uppercase Latin letters $I,J,K,\dots=0,1,2,3$ are used as ``internal indices'' for algebraic structure. When split into temporal and spatial parts, the internal indices take the form $I\rightarrow(0,i)$, where the lowercase Latin letters $i,j,k,\dots=1,2,3$ are used for the spatial indices.
On the other hand, the lowercase Greek letters $\alpha,\beta,\dots,\mu,\nu,\dots$ are used as ``external indices'' for spacetime coordinates. When split into temporal and spatial parts, the external indices take the form $\mu\equiv x^I\equiv(x^0,x^i) \equiv(x^0,x^1,x^2,x^3)$.
The spacetime signature is $(-,+,+,+)$.
The metric of a flat Minkowski spacetime is given by $\eta_{\mu\nu}=\mathrm{diag}(1,-1,-1,-1)$, and the metric of a Minkowskian internal space is given by $\eta_{IJ}=\mathrm{diag}(1,-1,-1,-1)$.
The reduced Planck constant and the speed of light are set to be unity, i.e., $\hbar=1$ and $c=1$.

\section{Tetrad formalism of linearized gravity}\label{sec:tetrad formalism}
Before studying the interaction of a hydrogen-like atom with gravitational waves, we first formulate the tetrad formalism \cite{carroll2019spacetime} of the linearized theory of gravity \cite{carroll2019spacetime, misner1973gravitation}, whereby the gravitational coupling can be readily prescribed in the Dirac equation.

In the tetrad formalism for a 4-dimensional spacetime manifold, we first choose a local basis for each point, i.e., a set of $4$ independent vector fields called the \emph{tetrad} fields:
\begin{equation}
e_{I} \equiv {e_I}^\mu \partial_\mu, \quad \text{for}\ I=0,1,2,3.
\end{equation}
Dually, a set of 4 independent 1-forms called the \emph{cotetrad} fields is given by
\begin{equation}
e^I \equiv {e^I}_\mu dx^\mu, \quad \text{for}\ I=0,1,2,3,
\end{equation}
such that
\begin{equation}
e^I(e_J) = \delta^I_J,
\end{equation}
which implies that ${e_I}^\mu$ and ${e^I}_\mu$ as $4\times4$ matrices are inverse to each other, i.e.,
\begin{subequations}\label{dual relations}
\begin{eqnarray}
{e_I}^\mu {e^J}_\mu &=& \delta_I^J, \\
{e_I}^\mu {e^I}_\nu &=& \delta_\nu^\mu.
\end{eqnarray}
\end{subequations}
The cotetrad $e^I$, which in a sense is the square root of the metric, gives rise to the metric via
\begin{equation}
g_{\mu\nu}= {e^I}_\mu {e^J}_\nu\eta_{IJ}.
\end{equation}
Correspondingly, by the fact that $g^{\mu\nu}$ and $\eta^{IJ}$ are the inverse matrices of $g_{\mu\nu}$ and $\eta_{IJ}$ respectively, it follow
\begin{equation}
g^{\mu\nu}= {e_I}^\mu {e_J}^\nu\eta^{IJ}.
\end{equation}
Consequently, we can define $e_{I\nu}$ and $e^{I\nu}$ such that
\begin{subequations}
\begin{eqnarray}
e_{I\nu} &:=& {e_I}^\mu g_{\mu\nu} = \eta_{IJ}{e^J}_\nu, \\
e^{I\nu} &:=& {e^I}_\mu g^{\mu\nu} = \eta^{IJ} {e_J}^\nu.
\end{eqnarray}
\end{subequations}

In the weak-field limit of gravity on top of a flat Minkowski background, the spacetime metric can be decomposed into
\begin{subequations}\label{eta and h}
\begin{eqnarray}
g_{\mu\nu} &=& \eta_{\mu\nu}+h_{\mu\nu},\\
g^{\mu\nu} &=& \eta^{\mu\nu}-h^{\mu\nu},
\end{eqnarray}
\end{subequations}
where $\abs{h_{\mu\nu}}\ll1$ is the first-order perturbation of gravity upon the flat background metric $\eta_{\mu\nu}$.
It follows from $g_{\mu\alpha}g^{\alpha\nu}=\delta_\mu^\nu$ that
\begin{equation}
h^{\mu\nu} = \eta^{\mu\alpha}\eta^{\nu\beta}h_{\alpha\beta}.
\end{equation}
In accordance with \eqref{eta and h}, we decompose the cotetrad and tetrad as
\begin{subequations}\label{e and b}
\begin{eqnarray}
{e^I}_\mu &=& {\underline{e}^I}_\mu + {b^I}_\mu,\\
{e_I}^\mu &=& {\underline{e}_I}^\mu - {b_I}^\mu,
\end{eqnarray}
\end{subequations}
where $\underline{e}^I$ and $\underline{e}_I$ are the cotetrad and tetrad for the flat background, i.e., $\eta_{\mu\nu}= {\underline{e}^I}_\mu {\underline{e}^J}_\nu\eta_{IJ}$ and $\eta^{\mu\nu}= {\underline{e}_I}^\mu {\underline{e}_J}^\nu\eta^{IJ}$, and $b^I$ and $b_I$ are assumed to be $O(\abs{h_{\mu\nu}})$.
Substituting \eqref{e and b} into \eqref{eta and h} and ignoring any terms higher than $O(\abs{h_{\mu\nu}})$, we obtain
\begin{subequations}\label{h and b}
\begin{eqnarray}
\label{h and b a}
h_{\mu\nu} &=& \eta_{IJ} \big({\underline{e}^I}_\mu {b^J}_\nu + {\underline{e}^I}_\nu {b^J}_\mu\big),\\
\label{h and b b}
h^{\mu\nu} &=& \eta^{IJ} \big({\underline{e}_I}^\mu {b_J}^\nu + {\underline{e}_I}^\nu {b_J}^\mu\big).
\end{eqnarray}
\end{subequations}
Meanwhile, substituting \eqref{e and b} into \eqref{dual relations} and ignoring the terms higher than $O(h_{\mu\nu})$, we arrive at
\begin{subequations}\label{e0 and b}
\begin{eqnarray}
{\underline{e}_I}^\mu {b^J}_\mu &=& {\underline{e}^J}_\mu {b_I}^\mu,\\
{\underline{e}_I}^\mu {b^I}_\nu &=& {\underline{e}^I}_\nu {b_I}^\mu,
\end{eqnarray}
\end{subequations}
which enables us to define $b^{I\nu}$ and $b_{I\nu}$ such that
\begin{subequations}\label{eta and b}
\begin{eqnarray}
b^{I\nu} &:=& {b^I}_\mu \eta^{\mu\nu} = \eta^{IJ} {b_J}^\nu,\\
b_{I\nu} &:=& {b_I}^\mu \eta_{\mu\nu} = \eta_{IJ} {b^J}_\nu.
\end{eqnarray}
\end{subequations}
Multiplying ${\underline{e}_K}^\mu$ on both sides of \eqref{h and b a} and ${\underline{e}^K}_\mu$ on both sides of \eqref{h and b b} and using \eqref{e0 and b} and \eqref{eta and b}, we obtain
\begin{subequations}\label{b in h}
\begin{eqnarray}
{b^I}_\mu &=& \frac{1}{2}\eta^{IJ} {\underline{e}_J}^\nu h_{\mu\nu} \equiv \frac{1}{2}\,\underline{e}^{I\nu}h_{\mu\nu},\\
{b_I}^\mu &=& \frac{1}{2}\eta_{IJ} {\underline{e}^J}_\nu h^{\mu\nu} \equiv
\frac{1}{2}\,\underline{e}_{I\nu}h^{\mu\nu}.
\end{eqnarray}
\end{subequations}
Given a fixed (co)tetrad field $\underline{e}^I$ and $\underline{e}_I$, the equation \eqref{b in h} translates the metric perturbation $h_{\mu\nu}$ and $h^{\mu\nu}$ into the (co)tetrad perturbation $b^I$ and $b_I$. Our beginning assumption that $b^I$ and $b_I$ are of $O(\abs{h_{\mu\nu}})$ is now consistently verified. The factor $1/2$ arises as expected, since the (co)tetrad can be intuitively understood as the square root of the metric.

Up to the order of $O(\abs{h_{\mu\nu}}^2)$, the Levi-Civita connection in terms of $h_{\mu\nu}$ is given by
\begin{equation}\label{Gamma in h}
{\Gamma^\alpha}_{\mu\nu}=\frac{1}{2} \eta^{\alpha\beta} (h_{\beta\nu,\mu}+h_{\mu\beta,\nu}-h_{\mu\nu,\beta}).
\end{equation}
The corresponding torsion-free spin connection is given by
\begin{eqnarray}\label{omega in h}
{\omega^{IJ}}_\mu
&:=& {e^I}_\alpha {\Gamma^\alpha}_{\mu\nu} e^{J\nu} -  e^{J\nu}\partial_\mu {e^I}_\nu \nonumber\\
&\approx& {\underline{e}^I}_\alpha {\Gamma^\alpha}_{\mu\nu} \underline{e}^{J\nu} -  \underline{e}^{J\nu}\, \partial_\mu {b^I}_\nu \nonumber\\
&=& \frac{1}{2}\underline{e}^{I\alpha}\underline{e}^{J\beta}
\left(h_{\mu\alpha,\beta}-h_{\mu\beta,\alpha}\right),
\end{eqnarray}
where \eqref{b in h} and \eqref{Gamma in h} have been used.

In the weak-field limit, it is natural to choose $\mu$ to be the ``flat'' Minkowski coordinates, i.e., $\mu\equiv x^I\equiv (x^0,x^1,x^2,x^3)=(t,x,y,z)$. Correspondingly, we can simply choose ${\underline{e}^I}_\mu$ to be aligned with the flat coordinates, i.e.,
\begin{subequations}\label{flat e}
\begin{eqnarray}
{\underline{e}^I}_\mu &\equiv& {\underline{e}^I}_{x^J}=\delta^I_J, \\
{\underline{e}_I}^\mu &\equiv& {\underline{e}_I}^{x^J}=\delta_I^J.
\end{eqnarray}
\end{subequations}
Furthermore, for simplicity, we consider a monochromatic gravitational wave prorogating in the $z$ direction. The field $h_{\mu\nu}$ of it in the transverse-traceless (TT) gauge takes the form \cite{misner1973gravitation}
\begin{equation}\label{h in TT}
h_{\mu\nu}(t,x,y,z)
          =\left( \begin{array}{cccc}
                    0 & 0 & 0 & 0 \\
                    0 & h_+ & h_\times & 0 \\
                    0 & h_\times & -h_+ & 0 \\
                    0 & 0 & 0 & 0
                  \end{array}
           \right)
          =\left( \begin{array}{cccc}
                    0 & 0 & 0 & 0 \\
                    0 & h^0_+ & h^0_\times & 0 \\
                    0 & h^0_\times & -h^0_+ & 0 \\
                    0 & 0 & 0 & 0
                  \end{array}
           \right)
           e^{i(kz-\omega t)}
          +\mathrm{c.c.},
\end{equation}
where $\omega = \abs{k}$ and the constants $h^0_+$ and $h^0_\times$ are the amplitudes of the ``plus'' and ``cross'' polarizations respectively.
With \eqref{flat e} and \eqref{h in TT}, it follows from \eqref{b in h} that
\begin{subequations}\label{b in GW}
\begin{eqnarray}
{b^1}_x &=& -{b^2}_y = -\frac{1}{2} h_+,\\
{b^1}_y &=& {b^2}_x  = -\frac{1}{2} h_\times,\\
{b^I}_\mu &=& 0\quad \text{otherwise},
\end{eqnarray}
\end{subequations}
and from \eqref{omega in h} that
\begin{subequations}\label{omega in GW}
\begin{eqnarray}
{\omega^{13}}_x &=& -{\omega^{31}}_x = -{\omega^{23}}_y = {\omega^{32}}_y = \frac{1}{2}\partial_z h_+,\\
{\omega^{23}}_x &=& -{\omega^{32}}_x = {\omega^{13}}_y = -{\omega^{31}}_y = \frac{1}{2}\partial_z h_\times,\\
{\omega^{10}}_x &=& -{\omega^{01}}_x = -{\omega^{20}}_y = {\omega^{02}}_y = -\frac{1}{2}\partial_t h_+,\\
{\omega^{20}}_x &=& -{\omega^{02}}_x = {\omega^{10}}_y = -{\omega^{01}}_y = -\frac{1}{2}\partial_t h_\times,\\
{\omega^{IJ}}_\mu &=& 0\quad \text{otherwise}.
\end{eqnarray}
\end{subequations}

\section{Hydrogen-like atoms in a gravitational wave}\label{sec:hydrogen-like atoms}
The relativistic quantum theory of a spin-$1/2$ particle with charge $e$ and mass $m_e$ subject to an external electromagnetic field in curved spacetime is described by the covariant Dirac equation \cite{birrell1984quantum}:
\begin{equation}\label{Dirac eq}
  \left(i\tilde{\gamma}^I {e_I}^\mu D_\mu - m_e\right) \ket{\psi}  = 0,
\end{equation}
where $\ket{\psi}=(\varphi,\chi)^T$ is the Dirac spinor composed of two 2-component Wyle spinor $\varphi$ and $\chi$, the $4\times4$ matrices $\tilde{\gamma}^I$ are the gamma matrices,\footnote{We decorate a notation with a tilde $\tilde{}$ to indicate that it is a $4\times4$ matrix for the Dirac spinor.} and $D_\mu$ is the covariant derivative.
The gamma matrices satisfy the anticommutation relation $\{\tilde{\gamma}^I,\tilde{\gamma}^J\}=2 \eta^{IJ} \tilde{I}$ and in the Dirac representation are given by $\tilde{\gamma}^0=\tilde{\beta}$ and $\tilde{\gamma}^i=\tilde{\beta}\tilde{\alpha}^i$ with
\begin{equation}
\tilde{\beta}=\left(\begin{array}{cc}
                      I & 0 \\
                      0 & -I
                    \end{array}\right),
\quad
\tilde{\alpha}^i=\left(\begin{array}{cc}
                      0 & \sigma^i \\
                      \sigma^i & 0
                    \end{array}\right),
\end{equation}
where $\sigma^i$ for $i=1,2,3$ denote the Pauli matrices.
The covariant derivative $D_\mu$ is given via \emph{minimal coupling} as
\begin{equation}
 D_\mu := \partial_\mu+ie A_\mu +\frac{i}{2} {\omega^{IJ}}_\mu\tilde{\sigma}_{IJ},
\end{equation}
where $A_\mu=(\phi,A_x,A_y,A_z)$ is the electromagnetic connection (potential), while $\omega_\mu$ is the spin connection with the non-abelian value $\tilde{\sigma}^{IJ}:=\frac{i}{2}[\tilde{\gamma}^I,\tilde{\gamma}^J]$.

In a hydrogen-like atom, because the nucleus is assumed to be much more massive than the electron, we can consider only the effect upon the electron due to the electromagnetic field sourced by the nucleus and ignore the back-reaction of the electron upon the nucleus.
That is, in the Dirac equation for the electron in a hydrogen-like atom, the electromagnetic connection is given by the Coulomb potential, i.e.,
\begin{equation}\label{Coulomb potential}
A_\mu(t,x,y,z) \equiv (\phi,A_x,A_y,A_z) = \left(-\frac{Ze}{r},0,0,0\right),
\end{equation}
where the nucleus is assumed to have charge $-Ze$ with $Z$ being the atomic number.
When this hydrogen-like atom is subject to an external gravitational field, there are three consequent effects. Firstly, the flat-spacetime tetrad ${\underline{e}^I}_\mu$ is replaced by a non-flat one ${e^I}_\mu$. Secondly, the spin connection ${\omega^{IJ}}_\mu$ is nonzero, giving corrections via the covariant derivative $D_\mu$. Finally, the electromagnetic potential \eqref{Coulomb potential} is modified in the presence of a gravitational wave.

Heuristically, we can posit that the Coulomb potential is modified as a result of the distance $r$ between the electron and the nucleus being dynamically contracted or dilated by the gravitational wave.
That is, we still have $A_\mu=(\phi,0,0,0)$ but the Coulomb potential $\phi$ is altered into
\begin{eqnarray}\label{phi}
\phi(t,x,y,z;g_{\mu\nu}) &=& -\frac{Ze}{\sqrt{-g_{ij}x^i x^j}}
=-\frac{Ze}{r}\left(1+\frac{ h_+(x^2-y^2)+2h_\times xy }{2r^2} \right) + O(\abs{h_{\mu\nu}}^2) \nonumber\\
&=:& \phi^{(0)} + \phi^{(1)} + O(\abs{h_{\mu\nu}}^2)
\end{eqnarray}
where, $r\equiv (x^2+y^2+z^2)^{1/2}$, $\phi^{(0)}=-Ze/r$ is the original (unperturbed) Coulomb potential, and $\phi^{(1)}$ is the corresponding first-order correction.
This heuristic prescription provides an intuitive picture of how the Coulomb interaction is modified, but in fact it also agrees up to $O(\abs{h_{\mu\nu}}^2)$ with the result derived from a more fundamental ground in the long-wavelength limit (i.e.\ the wavelength $\lambda=2\pi/\abs{k}$ of the gravitational wave is assumed to be much larger than the atomic size).

In general relativity, Maxwell's equations in curved spacetime are given by \cite{birrell1984quantum}
\begin{subequations}
\begin{eqnarray}
\label{Maxwell a}
F_{\alpha\beta} &=& \partial_\alpha A_\beta - \partial_\beta A_\alpha, \\
\label{Maxwell b}
\mathcal{D}^{\mu\nu} &=& \sqrt{-g} g^{\mu\alpha}g^{\nu\beta}F_{\alpha\beta},\\
\label{Maxwell c}
\mathcal{J}^\mu &=& \partial_\nu \mathcal{D}^{\mu\nu},
\end{eqnarray}
\end{subequations}
where the two-form $F_{\alpha\beta}$ is the electromagnetic field, the weight-1 tensor density $\mathcal{D}^{\mu\nu}$ is the electric displacement field and the auxiliary magnetic field, and the weight-1 vector density $\mathcal{J}^\mu$ is the electric current density. Note that these equations are covariant despite the use of ordinary partial derivatives instead of covariant derivatives.
Since we ignore the back-reaction upon the nucleus, the nucleus follows a geodesic and thus stays at the origin even in the presence of a gravitational wave.
By assuming that the nucleus is a point particle, the current density $\mathcal{J}^\mu$ is given by a point charge $-Ze$ sitting at the origin. The explicit formula of $\mathcal{J}^\mu$ is difficult to obtain directly, because its singular nature makes the weak-field and long-wavelength limits rather cumbersome. By contrast, the formula of $\mathcal{D}^{\mu\nu}$ is much easier. In the weak-field and long-wavelength limit, we can safely neglect any electromagnetic radiation from the nucleus due to the dynamical deformation of its charge distribution induced by the gravitational wave, and thus, at any instantaneous moment $t$, the tensor density $\mathcal{D}^{\mu\nu}$ is just the electric displacement field emanating from a point charge at origin.
At any moment $t$, the metric given by \eqref{eta and h} with \eqref{h in TT} can be diagonalized to yield the ``orthonormal coordinates'' $\bar{x}^I\equiv (\bar{t},\bar{x},\bar{y},\bar{z})$ as a linear transform from $(t,x,y,z)$ such that $g_{\bar{x}^\mu\bar{x}^\nu}=\eta_{\mu\nu}$. Obviously, we have $\bar{t}=t$ and $\bar{z}=z$, and more precisely, $\bar{x}=\bar{x}(x,y;h_+,h_\times)$ and $\bar{y}=\bar{y}(x,y;h_+,h_\times)$ are linear transforms from $x$ and $y$ with linear coefficients depending on $h_+(t,z)$ and $h_\times(t,z)$.

In the orthonormal coordinates, the electric field of a point charge simply takes the familiar inverse-square law, i.e.,
\begin{subequations}
\begin{eqnarray}
\mathcal{D}^{\bar{t}\bar{x}^i}\partial_{\bar{x}^i} &=& - \mathcal{D}^{\bar{x}^i\bar{t}}\partial_{\bar{x}^i}\approx Ze\frac{\bar{x}^i\partial_{\bar{x}^i}}
{(\bar{x}^2+\bar{y}^2+\bar{z}^2)^{3/2}}, \\
\mathcal{D}^{\bar{t}\bar{t}} &=& 0, \qquad \mathcal{D}^{\bar{x}^i\bar{x}^j} \approx 0.
\end{eqnarray}
\end{subequations}
Since $(\bar{x},\bar{y},\bar{z})$ is a linear transform from $(x,y,z)$, we have $\bar{x}^i\partial_{\bar{x}^i}=x^i\partial_{x^i}$. Furthermore, $\bar{x}^2+\bar{y}^2+\bar{z}^2\equiv -g_{\bar{x}^i\bar{x}^j}\bar{x}^i\bar{x}^j = -g_{x^ix^j}x^ix^j$.
Therefore, in the coordinates $(t,x,y,z)$, the tensor density $\mathcal{D}^{\mu\nu}$ reads as
\begin{subequations}
\begin{eqnarray}
\mathcal{D}^{tx^i}\partial_{x^i} &=& - \mathcal{D}^{x^it}\partial_{x^i}\approx Ze\frac{x^i\partial_{x^i}}
{\left((1+h_+)x^2+(1-h_+)y^2+2h_\times xy+z^2\right)^{3/2}}, \nonumber\\
&=& Ze \left[\frac{1}{r^3}- \frac{3x\left(h_+(x^2-y^2)+2h_\times xy\right)}{2r^5}\right] x^i\partial_{x^i}
+ O(\abs{h_{\mu\nu}}^2), \\
\mathcal{D}^{tt} &=& 0, \qquad \mathcal{D}^{x^ix^j} \approx 0.
\end{eqnarray}
\end{subequations}
By applying \eqref{Maxwell b}, which implies $F_{\alpha\beta}=g_{\alpha\mu}g_{\beta\nu}\mathcal{D^{\mu\nu}}/\sqrt{-g}$, and substituting \eqref{eta and h} with \eqref{h in TT} for $g_{\mu\nu}$, we then obtain
\begin{subequations}
\begin{eqnarray}
F_{tx} &=& -F_{xt} \approx -Ze\left[\frac{x}{r^3} + \frac{h_+x+h_\times y}{r^3} - \frac{3x\left(h_+(x^2-y^2)+2h_\times xy\right)}{2r^5}\right], \\
F_{ty} &=& -F_{yt} \approx -Ze\left[\frac{y}{r^3} + \frac{h_\times x-h_+ y}{r^3} - \frac{3y\left(h_+(x^2-y^2)+2h_\times xy\right)}{2r^5}\right], \\
F_{tz} &=& -F_{zt} \approx -Ze\left[\frac{z}{r^3} - \frac{3z\left(h_+(x^2-y^2)+2h_\times xy\right)}{2r^5}\right], \\
F_{\mu\nu} &=& -F_{\nu\mu} \approx 0, \qquad \text{otherwise}.
\end{eqnarray}
\end{subequations}
According to \eqref{Maxwell a}, it is straightforward to show that the corresponding potential $A_\mu$ is given by $A_\mu=(\phi, 0, 0, 0)$ with $\phi$ being the one given in \eqref{phi}.

Now, take into account all the three effects due to a gravitational wave. In the weak-field and long-wavelength limit, choosing the Minkowski coordinate $\mu\equiv x^I=(t,x,y,z)$ and adopting \eqref{flat e}, we have
\begin{eqnarray}
i\tilde{\gamma}^I {e^\mu}_I D_\mu
&=& i\tilde{\gamma}^I \left(\delta_I^{x^J}+{b_I}^{x^J}\right)
\left(\partial_{x^J}+ie A_{x^J} +\frac{i}{2} {\omega^{IK}}_{x^J}\tilde{\sigma}_{IK}\right) + O(\abs{h_{\mu\nu}}^2)\nonumber\\
&=& i\tilde{\gamma}^I\left(\partial_{x^I}+ieA_{x^I}\right)
+i\tilde{\gamma}^I{b_I}^\mu\left(\partial_\mu+ieA_\mu\right)
-\frac{1}{2}\tilde{\gamma}^I{\omega^{JK}}_{x^I}\tilde{\sigma}_{JK} + O(\abs{h_{\mu\nu}}^2),
\end{eqnarray}
which, with \eqref{b in GW} and \eqref{omega in GW}, is further simplified into [The leading factor is corrected as the correction of a factor of $-1/2$ in \eqref{b in GW}.]
\begin{eqnarray}\label{Dmu term}
i\tilde{\gamma}^I {e^\mu}_I D_\mu
&\approx& i\tilde{\gamma}^0\left(\partial_t+ie\phi^{(0)}+ie\phi^{(1)}\right)
+i\tilde{\gamma}^i\partial_{x^i}\nonumber\\
&& \mbox{}
-\frac{i}{2}\tilde{\gamma}^1\left(h_+\partial_x+h_\times\partial_y\right)
-\frac{i}{2}\tilde{\gamma}^2\left(h_\times\partial_x-h_+\partial_y\right) \nonumber\\
&& \mbox{}
-\frac{1}{2}\partial_zh_+
\cancelto{0}{\left(\tilde{\gamma}^1\tilde{\sigma}_{13}-\tilde{\gamma}^2\tilde{\sigma}_{23}\right)}
-\frac{1}{2}\partial_zh_\times
\cancelto{0}{\left(\tilde{\gamma}^1\tilde{\sigma}_{23}-\tilde{\gamma}^2\tilde{\sigma}_{31}\right)}\nonumber\\
&& \mbox{}
+\frac{1}{2}\partial_th_+
\cancelto{0}{\left(\tilde{\gamma}^1\tilde{\sigma}_{10}-\tilde{\gamma}^2\tilde{\sigma}_{20}\right)}
+\frac{1}{2}\partial_th_\times
\cancelto{0}{\left(\tilde{\gamma}^1\tilde{\sigma}_{20}-\tilde{\gamma}^2\tilde{\sigma}_{01}\right)},
\end{eqnarray}
where the last line vanishes as it is easy to show $\tilde{\gamma}^1\tilde{\sigma}_{13}-\tilde{\gamma}^2\tilde{\sigma}_{23} = \tilde{\gamma}^1\tilde{\sigma}_{23}-\tilde{\gamma}^2\tilde{\sigma}_{31} = \tilde{\gamma}^1\tilde{\sigma}_{10}-\tilde{\gamma}^2\tilde{\sigma}_{20} = \tilde{\gamma}^1\tilde{\sigma}_{20}-\tilde{\gamma}^2\tilde{\sigma}_{01} =0$.
By substituting \eqref{Dmu term} into \eqref{Dirac eq} and recasting the Dirac equation into the Hamiltonian form
\begin{equation}
i\frac{\partial}{\partial t}\ket{\psi }=H\ket{\psi},
\end{equation}
the Hamiltonian is given by
\begin{equation}\label{H}
H = H_0 + H^{(1)}_\phi + H^{(1)}_e + O(\abs{h_{\mu\nu}}^2)
\end{equation}
with the three terms as discussed below.

The first term is given by
\begin{eqnarray}
H_0 &=& \tilde{\gamma}^0m_e -i\tilde{\gamma}^0\left(ie\tilde{\gamma}^0\phi^{(0)}+\tilde{\gamma}^i\partial_{x^i}\right) \nonumber\\
&=& \tilde{\beta}m_e + e\phi^{(0)} + \tilde{\alpha}^ip_{x^i},
\end{eqnarray}
where $p_{x^i}\equiv-i\partial_{x^i}\equiv-i(\partial_x,\partial_y,\partial_z)$.
This is the original (unperturbed) Hamiltonian without any external gravitational field.
The second term is given by
\begin{equation}\label{H1 phi}
H^{(1)}_\phi = e\phi^{(1)} = \frac{Ze^2}{2r^3}\left(h_+(x^2-y^2)+2h_\times xy\right),
\end{equation}
which is the first-order gravitational-wave correction upon the Coulomb potential.
The third term is given by
\begin{eqnarray}\label{H1 e}
H^{(1)}_e &=&
\frac{i}{2}\tilde{\gamma}^0\tilde{\gamma}^1\left(h_+\partial_x+h_\times\partial_y\right)
\frac{i}{2}\tilde{\gamma}^0\tilde{\gamma}^2\left(h_\times\partial_x-h_+\partial_y\right) \nonumber\\
&=&
\frac{i}{2}\tilde{\alpha}^1\left(h_+\partial_x+h_\times\partial_y\right)
\frac{i}{2}\tilde{\alpha}^2\left(h_\times\partial_x-h_+\partial_y\right) \nonumber\\
&=& \frac{i}{2}\sigma^1\otimes
\left(
  \begin{array}{cc}
    0 & (h_+-ih_\times)(\partial_x+i\partial_y) \\
    (h_++ih_\times)(\partial_x-i\partial_y) & 0 \\
  \end{array}
\right),
\end{eqnarray}
which is the first-order gravitational correction directly via the tetrad.
Also note that, as the last line in \eqref{Dmu term} vanishes identically, the gravitational wave does not give rise to any first-order correction via the spin connection.
In the next section, we investigate the effects of these three terms in depth.

\section{Energy levels and atomic transitions}\label{sec:atomic transitions}
In this section, we first present the exact solutions to the Dirac equation with only the unperturbed Hamiltonian $H_0$ \cite{strange1998relativistic,greiner2000relativistic}. Later, we consider the Coulomb potential correction term $H^{(1)}_\phi$ and the tetrad correction term $H^{(1)}_e$ separately. In the presence of $H^{(1)}_\phi$ and $H^{(1)}_e$, the Dirac equation can no longer be exactly solved. Fortunately, we can apply the first-order perturbation theory to study the atomic transitions induced by $H^{(1)}_\phi$ and $H^{(1)}_e$.

\subsection{Original Hamiltonian $H_0$}
The unperturbed Hamiltonian $H_0$ describes a charged Dirac particle in a Coulomb potential, which can be exactly solved (see e.g.\ \cite{strange1998relativistic,greiner2000relativistic} and especially Exercise 9.6 in \cite{greiner2000relativistic}).
In a spherically symmetric potential, i.e., $\phi^{(0)}(x,y,z)=\phi^{(0)}(r)$, the Dirac Hamiltonian $H_0=\tilde{\beta}m + e\phi^{(0)} + \alpha^ip_i$ commute with the total angular momentum $J^i:=L^i+S^i\equiv L^i+\frac{1}{2}\tilde{\sigma}^i$ for $i=1,2,3$, where $L^i:=-i(y\partial_z-z\partial_y,z\partial_x-x\partial_z,x\partial_y-y\partial_x)$ is the orbital angular momentum, and $S^i\equiv\frac{1}{2}\tilde{\sigma}^i$ (or $S^i\equiv\frac{\hbar}{2}\tilde{\sigma}^i$ with $\hbar$ explicitly shown) is the spin angular momentum with
\begin{equation}
\tilde{\sigma}^i=\left(\begin{array}{cc}
                      \sigma^i & 0 \\
                      0 &  \sigma^i
                    \end{array}\right).
\end{equation}
Therefore, the eigenfunctions of $H_0$ can be specified as also eigenfunctions of the operators $J^2:=(J^1)^2+(J^2)^2+(J^3)^2$ and $J^3$ with eigenvalues $j(j+1)$ and $m_j$ respectively.
More precisely \cite{greiner2000relativistic}, the eigenfunctions of $H_0$ are given by the Dirac spinors
\begin{equation}\label{psi jmj}
\psi_{jm_j} =
\left(
 \begin{array}{c}
  \varphi_{jlm_j} \\
  \chi_{jl'm_j} \\
 \end{array}
\right)
=
\left(
  \begin{array}{c}
    ig(r) \Omega_{jlm_j}(\theta,\phi)\\
    -f(r) \Omega_{jl'm_j}(\theta,\phi)\\
  \end{array}
\right),
\end{equation}
where $j=l+1/2=l'-1/2$ and
\begin{subequations}
\begin{eqnarray}
\Omega_{j=l+\frac{1}{2},l,m_j} &=&
\left(
 \begin{array}{c}
  \sqrt{\frac{j+m_j}{2j}}\,Y_{l,m_j-\frac{1}{2}} \\
  \sqrt{\frac{j-m_j}{2j}}\,Y_{l,m_j+\frac{1}{2}} \\
 \end{array}
\right),
\\
\Omega_{j=l-\frac{1}{2},l,m_j} &=&
\left(
 \begin{array}{c}
  -\sqrt{\frac{j-m_j+1}{2j+2}}\,Y_{l,m_j-\frac{1}{2}} \\
  \sqrt{\frac{j+m_j+1}{2j+2}}\,Y_{l,m_j+\frac{1}{2}} \\
 \end{array}
\right).
\end{eqnarray}
\end{subequations}
The functions $g(r)$ and $f(r)$ are to be determined by solving the radial differential equation that depends on the exact form of $\phi^{(0)}(r)$.

In the case of a Coulomb potential $\phi^{(0)}(r)=-Ze/r$, solving the radial equation yields the discrete energy eigenvalues for the bound states specified by a \emph{principle quantum number} $n=1,2,3,\dots$ along with $j$ as
\begin{equation}\label{energy eigenvalue}
E = m_e
\left[
1+\frac{(Z\alpha)^2}{\left[n-j-\frac{1}{2}+\left[(j+\frac{1}{2})^2-(Z\alpha)^2\right]^{1/2}\right]^2}
\right]^{-1/2},
\end{equation}
where $\alpha\equiv e^2\approx 1/137$ (or $\alpha\equiv e^2/\hbar c$ with $\hbar$ and $c$ shown explicitly) is the fine-structure constant.
The corresponding eigenstates are symbolically denoted as $\ket{n,j,m_j}$.
For $Z\alpha\ll1$ (i.e., $Z\ll137$), the energy can be approximated as
\begin{equation}
E \approx m_e -\frac{m_e(Z\alpha)^2}{2n^2} -\frac{(Z\alpha)^4}{2n^3}\left(\frac{1}{j+\frac{1}{2}}-\frac{3}{4n}\right),
\end{equation}
where the first term is the rest energy, the second term is the celebrated Bohr-model energy, and the third term is the familiar fine-structure correction. The Dirac equation automatically gives rise to the fine-structure correction, which accounts for both the spin-orbit coupling and the relativistic correction.

In the absence of external gravitational fields, the Dirac equation coupled with the electromagnetic potential $A_\mu\equiv(\phi,A_x,A_y,A_z)$ gives the energy eigenvalue equation
\begin{equation}
\left(
  \begin{array}{cc}
    E-m_e-e\phi & -\sigma^i\pi_i \\
    -\sigma^i\pi_i & E+m_e-e\phi \\
  \end{array}
\right)
\left(
  \begin{array}{c}
    \varphi \\
    \chi \\
  \end{array}
\right)
=0,
\end{equation}
which leads to
\begin{equation}
\chi = \left(\frac{\sigma^i\pi_i}{E+m_e-e\phi}\right)\varphi,
\end{equation}
and $(E-e\phi)^2=m_e^2+\Pi^2$,
where $\Pi_{x^i}:=p_{x^i}-eA_{x^i}$ is the kinematic momentum and $\Pi^2:=(\Pi_x)^2+(\Pi_y)^2+(\Pi_z)^2$.
If we assume the nonrelativistic limit that the energy apart from the rest energy is much smaller than the rest energy, i.e., $E_s:=E-m_e\ll m_e$, and the electrostatic energy $e\phi$ is also much smaller than the rest energy, i.e., $\abs{e\phi}\ll m_e$, it leads to the Schr\"{o}dinger--Pauli approximation
\begin{equation}\label{Pauli approx}
\chi \approx \left(\frac{\sigma^i\pi_{x^i}}{2m_e}\right)\varphi,
\end{equation}
which implies that the part $\varphi$ dominates over $\chi$ as $\abs{\chi/\varphi}\approx v/2\ll1$, where $v$ is the velocity.
In the case of a hydrogen-like atom with $e\phi=-Z\alpha/r$, it follows from the virial theorem and \eqref{energy eigenvalue} that the Schr\"{o}dinger--Pauli approximation is valid for $Z\alpha\ll1$.

\subsection{Coulomb potential correction term $H^{(1)}_\phi$}
The perturbation term $H^{(1)}_\phi$ given by \eqref{H1 phi} with \eqref{h in TT} takes the explicit form
\begin{eqnarray}
H^{(1)}_\phi
&=&\frac{Z\alpha}{2r^3}
\left[\left(h^0_+(x^2-y^2)+2h^0_\times xy\right)e^{i(kz-\omega t)}
+ \left(h^{0*}_+(x^2-y^2)+2h^{0*}_\times xy\right)e^{-i(kz-\omega t)}
\right] \nonumber\\
&=:& H^{(+)}_\phi e^{-i\omega t} + H^{(-)}_\phi e^{i\omega t},
\end{eqnarray}
where $H^{(+)}_\phi$ and $H^{(-)}_\phi$ are the positive- and negative-frequency parts respectively.
According to the first-order perturbation theory for the periodic perturbation \cite{shankar1994principles,sakurai2021modern}, the average transition rate from the initial state $\ket{i}=\ket{njm_j}$ to the final state $\ket{f}=\ket{n'j'm'_j}$ is
\begin{eqnarray}
R^{(\phi)}_{i\rightarrow f}&=& R^{(\phi,+)}_{i\rightarrow f} + R^{(\phi,-)}_{i\rightarrow f} \nonumber\\
&=& 2\pi\abs{\bra{f}H^{(+)}_\phi\ket{i}}^2 \delta(E_f-E_i-\omega)
+ 2\pi\abs{\bra{f}H^{(-)}_\phi\ket{i}}^2 \delta(E_f-E_i+\omega),
\end{eqnarray}
where $E_i$ and $E_f$ are given by \eqref{energy eigenvalue} and the Dirac delta functions indicate Fermi's golden rule.
In order for the transition rate to be appreciable, Fermi's golden rule dictates that
\begin{equation}
\omega=\abs{k}\approx\abs{E_f-E_i}
\sim O\left(\frac{m_e(Z\alpha)^2}{2n^2}\right).
\end{equation}
On the other hand, according to the Bohr model, the radius at the level $n$ of the hydrogen-like atom is
\begin{equation}
r_n = \frac{n^2}{m_eZ\alpha}.
\end{equation}
The ratio of the atomic size $a$ to the wavelength $\lambda=2\pi/\abs{k}$ of the gravitational wave is of the order
\begin{equation}
\frac{a}{\lambda} \sim O\left(\frac{Z\alpha}{4\pi}\right).
\end{equation}
Therefore, for $Z\alpha\ll1$, we can take the long-wavelength limit $a/\lambda\ll1$ and assume $e^{\pm ikz}\approx 1$.

Consequently,
\begin{subequations}
\begin{eqnarray}
\bra{f}H^{(+)}_\phi\ket{i} &=&
Z\alpha
\bra{f}
\frac{h^0_+(x^2-y^2)+2h^0_\times xy}{2r^3}
\ket{i} \nonumber\\
&=& Z\alpha\sqrt{\frac{4\pi}{15}}
\left[
h_L^0\,\bra{f}\frac{Y_2^2}{r}\ket{i}
+
h_R^0\,\bra{f}\frac{Y_2^{-2}}{r}\ket{i}
\right], \\
\bra{f}H^{(-)}_\phi\ket{i} &=&
Z\alpha
\bra{f}
\frac{(h^0_+)^*(x^2-y^2)+2(h^0_\times)^* xy}{2r^3}
\ket{i} \nonumber\\
&=& Z\alpha\sqrt{\frac{4\pi}{15}}
\left[
(h_R^0)^*\bra{f}\frac{Y_2^2}{r}\ket{i}
+
(h_L^0)^*\bra{f}\frac{Y_2^{-2}}{r}\ket{i}
\right],
\end{eqnarray}
\end{subequations}
where we have expressed $(x^2-y^2)/r^2=\sqrt{8\pi/15}\left(Y_2^2+Y_2^{-2}\right)$ and $(2ixy)/r^2=\sqrt{8\pi/15}\left(Y_2^2-Y_2^{-2}\right)$ in terms of spherical harmonics and defined the amplitudes of the right- and left-handed circularly polarized modes as
\begin{subequations}
\begin{eqnarray}
h_R^0 &=& \frac{1}{\sqrt{2}}(h_+^0+ih_\times^0),\\
h_L^0 &=& \frac{1}{\sqrt{2}}(h_+^0-ih_\times^0).
\end{eqnarray}
\end{subequations}
According to the Wigner--Eckart theorem \cite{shankar1994principles,sakurai2021modern}, we have
\begin{equation}\label{Wigner-Eckart}
\bra{f}\frac{Y_2^{\pm2}}{r}\ket{i}
\equiv \bra{n'j'm'_j}\frac{Y_2^{\pm2}}{r}\ket{njm_j}
=\bra{n'j'}|T^{(k=2)}_\phi|\ket{nj}
\inner{j'm'_j}{2,\pm2;j,m_j},
\end{equation}
where $\bra{n'j'}|T^{(k=2)}_\phi|\ket{nj}$ is the corresponding \emph{reduced matrix element}, and $\inner{j'm'_j}{2,\pm2;j,m_j}$ are the Clebsch--Gordan coefficients, which vanish identically unless $j'\in\left\{\abs{j-2}, \abs{j-2}+1, \dots,j+2\right\}$ and $m'_j=m_j\pm2$.

Therefore, a gravitational wave will induce the excitation from a state $\ket{njm_j}$ of lower energy $E_{nj}$ to another state $\ket{n'j'm'_j}$ of higher energy $E_{n'j'}$ if $\omega\approx E_{n'j'}-E_{nj}$ or the de-excitation from higher energy $E_{nj}$ to lower energy $E_{n'j'}$ if $\omega\approx E_{nj}-E_{n'j'}$, following the \emph{selection rule} according to the Clebsch--Gordan coefficients.
The selection rule dictates that, for the case of $E_{n'j'}>E_{nj}$, the transition induced by the left-handed mode $h_L^0$ renders $j'\in\left\{\abs{j-2}, \abs{j-2}+1, \dots,j+2\right\}$ and $m'_j=m_j+2$, whereas that induced by the right-handed mode $h_R^0$ renders $j'\in\left\{\abs{j-2}, \abs{j-2}+1, \dots,j+2\right\}$ and $m'_j=m_j-2$. For the case of $E_{n'j'}<E_{nj}$, the transition induced by the left-handed mode $h_L^0$ renders $j'\in\left\{\abs{j-2}, \abs{j-2}+1, \dots,j+2\right\}$ and $m'_j=m_j-2$, whereas the transition induced by the right-handed mode $h_R^0$ renders $j'\in\left\{\abs{j-2}, \abs{j-2}+1, \dots,j+2\right\}$ and $m'_j=m_j+2$.

In the Schr\"{o}dinger--Pauli approximation, $\varphi_{jlm_j}$ dominates over $\chi_{jl'm_j}$ in \eqref{psi jmj} and the radial function $g(r)$ becomes very close to $R_{nl}(r)$ with $l=j-1/2$, which is the familiar radial function of the energy eigenstate for the nonrelativistic Schr\"{o}dinger equation of the hydrogen-like atom. Accordingly, the numerical value of the reduced matrix element can be numerically calculated using $R_{nl}(r)$, and its magnitude can be estimated as
\begin{equation}
\bra{n'j'}|T^{(k=2)}_\phi|\ket{nj}
=\int_0^\infty R^*_{n'l'}(r)\frac{1}{r}R_{nl}(r)r^2dr
\sim O\left(\frac{1}{r_n}\right) \sim O(m_eZ\alpha).
\end{equation}
Thus, the average transition rate is in the magnitude
\begin{equation}\label{R phi estimate}
R^{(\phi,\pm)}_{i\rightarrow f} \sim m_e^2(Z\alpha)^3 \big|h^0_{R/L}\big|^2\delta(E_f-E_i\mp\omega)
\end{equation}
up to the Clebsch--Gordan coefficients and a dimensionless factor of $O(1)$ depending on $n',j',n,j$.\footnote{Fermi's golden rule manifested as a delta function is a typical feature of a periodic perturbation. The singular nature of the delta function will disappear in real applications, as the transition amplitude has to be integrated over one way or another \cite{shankar1994principles}. As a simple example, if we consider a non-monochromatic gravitational wave, we have $h_{+/\times}(t,x,y,z)=\int dw \big(\mathpzc{h}^0_{+/\times}(\omega)e^{i(kx-\omega t)}+\mathrm{c.c.}\big)$, where the differential amplitudes $\mathpzc{h}^0_{+/\times}(\omega)$ with respect to $\omega$ are substituted for the constant amplitudes $h^0_{+/\times}$. Correspondingly, the total transition probability is given by $P^{(\phi,\pm)}_{i\rightarrow f} =\int_{-\infty}^\infty \dot{P}^{(\phi,\pm)}_{i\rightarrow f}dt \sim m_e^2(Z\alpha)^3\int d\omega \big|\mathpzc{h}^0_{R/L}(\omega)\big|^2 \delta(E_f-E_i\mp\omega)$, where $\mathpzc{h}^0_{R/L}=(\mathpzc{h}^0_+\pm i\mathpzc{h}^0_\times)/\sqrt{2}$.}

Even though we treat the gravitational wave purely as a classical field without any field quantization, the selection rule along with Fermi's golden rule strongly suggests that a gravitational wave carries \emph{gravitons} as massless spin-2 particles. The energy of a graviton is given by $\omega$ (or $\hbar\,\omega$ with $\hbar$ explicitly shown). As a massless spin-2 particle, the left- and right-handed modes are the two independent polarizations associated with the angular momentum $\pm2$ (or $\pm2\hbar$ with $\hbar$ explicitly shown) in the longitudinal (i.e.\ propagating) direction. The selection rule with Fermi's golden can be neatly understood in terms of the massless spin-2 gravitons. The average number flux of gravitons is proportional to the amplitude square of the gravitational wave, as $R^{(\phi,\pm)}_{i\rightarrow f} \propto |h^0_{R/L}|^2$ in \eqref{R phi estimate}.

\subsection{Tetrad correction term $H^{(1)}_e$}
The perturbation term $H^{(1)}_e$ given by \eqref{H1 e} with \eqref{h in TT} takes the explicit form
\begin{eqnarray}
H^{(1)}_e &=&
-\frac{1}{2\sqrt{2}}\,\sigma^1\otimes
\left(h_L^0\,p_+\sigma_+ + h_R^0\,p_-\sigma_-\right)
e^{i(kz-\omega t)} \nonumber\\
&&\mbox{}
-\frac{1}{2\sqrt{2}}\,\sigma^1\otimes
\left((h_R^0)^*p_+\sigma_+ + (h_L^0)^*p_-\sigma_-\right)
e^{-i(kz-\omega t)} \nonumber\\
&=:& H^{(+)}_e e^{-i\omega t} + H^{(-)}_e e^{i\omega t},
\end{eqnarray}
where $H^{(+)}_e$ and $H^{(-)}_e$ are the positive- and negative-frequency parts respectively, and we have defined
\begin{eqnarray}
\sigma_\pm &:=& \sigma^1 \pm i\sigma^2, \\
p_\pm &:=& p_x \pm i p_y.
\end{eqnarray}
According to the first-order perturbation theory, the transition rate from $\ket{i}=\ket{njm_j}$ to $\ket{f}=\ket{n'j'm'_j}$ is
\begin{eqnarray}
R^{(e)}_{i\rightarrow f}&=& R^{(e,+)}_{i\rightarrow f} + R^{(e,-)}_{i\rightarrow f} \nonumber\\
&=& 2\pi\abs{\bra{f}H^{(+)}_e\ket{i}}^2 \delta(E_f-E_i-\omega)
+ 2\pi\abs{\bra{f}H^{(-)}_e\ket{i}}^2 \delta(E_f-E_i+\omega),
\end{eqnarray}
where the Dirac delta functions again indicate Fermi's golden rule.

In the Dirac representation, $H^{(1)}_e$ is block off-diagonal. To simplify the matrix elements $\bra{f}H^{(\pm)}_e\ket{i}$, we take the Schr\"{o}dinger--Pauli approximation \eqref{Pauli approx} and let the Wyle spinor $\varphi\equiv\ket{\varphi_{njm_j}}$ be the eigenstate of the corresponding nonrelativistic Schr\"{o}dinger equation.
Taking the long-wavelength limit $e^{\pm ikz}\approx 1$ and also noting that $\pi_{x^i}=p_{x^i}$ in our case as $A_{x^i}=0$, we then have
\begin{subequations}
\begin{eqnarray}
\bra{f}H^{(+)}_e\ket{i} &=&
-\frac{1}{2\sqrt{2}\,m_e}
\bra{\varphi_{n'j'm'_j}}
\sigma^ip_{x^i}
\left(h_L^0\,p_+\sigma_+ + h_R^0\,p_-\sigma_-\right)
\ket{\varphi_{njm_j}}, \\
\bra{f}H^{(-)}_e\ket{i} &=&
-\frac{1}{2\sqrt{2}\,m_e}
\bra{\varphi_{n'j'm'_j}}
\sigma^ip_{x^i}
\left((h_R^0)^*p_+\sigma_+ + (h_L^0)^*p_-\sigma_-\right)
\ket{\varphi_{njm_j}}.
\end{eqnarray}
\end{subequations}
It is well known that the tensor product of two vectors can be decomposed into rank-0, rank-1, and rank-2 spherical tensors \cite{sakurai2021modern}. For $\bm{\sigma}\equiv(\sigma^1,\sigma^2,\sigma^3)$ and $\bm{p}\equiv(p_x,p_y,p_x)$, we have
\begin{subequations}
\begin{eqnarray}
T^{(0)}_0 &=& \frac{-\bm{\sigma}\cdot\bm{p}}{3} = \frac{-\sigma^ip_{x^i}}{3}, \\
T^{(1)}_q &=& \frac{(\bm{\sigma}\times\bm{p})^{(1)}_q}{i\sqrt{2}}, \\
T^{(2)}_{\pm2} &=& \sigma_{\pm1}p_{\pm1} = \frac{\sigma_\pm p_\pm}{2}, \\
T^{(2)}_{\pm1} &=& \frac{\sigma_{\pm1}p_{0}+\sigma_{0}p_{\pm1}}{\sqrt{2}}, \\
T^{(2)}_{0} &=& \frac{\sigma_{+1}p_{-1} +2\sigma_{0}p_{0} +\sigma_{-1}p_{+1}}{\sqrt{6}},
\end{eqnarray}
\end{subequations}
where $\sigma_{+1}:=-\sigma_+/\sqrt{2}$, $\sigma_{-1}:=\sigma_-/\sqrt{2}$, $\sigma_0:=\sigma^3$, $p_{+1}:=-p_+/\sqrt{2}$, $p_{-1}:=p_-/\sqrt{2}$, and $p_0:=p_z$.
In terms of the spherical tensors,
\begin{subequations}
\begin{eqnarray}
\bra{f}H^{(+)}_e\ket{i} &=&
\frac{3}{\sqrt{2}\,m_e}
\bra{\varphi_{n'j'm'_j}}
h_L^0\,T^{(0)}_0T^{(2)}_{2} + h_R^0\,T^{(0)}_0T^{(2)}_{-2}
\ket{\varphi_{njm_j}}, \\
\bra{f}H^{(-)}_e\ket{i} &=&
\frac{3}{\sqrt{2}\,m_e}
\bra{\varphi_{n'j'm'_j}}
(h_R^0)^*T^{(0)}_0T^{(2)}_{2} + (h_L^0)^*T^{(0)}_0T^{(2)}_{-2}
\ket{\varphi_{njm_j}}.
\end{eqnarray}
\end{subequations}
By the Wigner--Eckart theorem, we have
\begin{equation}
\bra{f}T^{(0)}_0T^{(2)}_{\pm2}\ket{i}
=\bra{n'j'}|T^{(k=2)}_e|\ket{nj}
\inner{j'm'_j}{2,\pm2;j,m_j},
\end{equation}
where $\bra{n'j'}|T^{(k=2)}_e|\ket{nj}$ is the corresponding \emph{reduced matrix element}, and the Clebsch--Gordan coefficients $\inner{j'm'_j}{2,\pm2;j,m_j}$ are exactly the same as in \eqref{Wigner-Eckart}.
The numerical value of the reduced matrix element can be numerically calculated using $\ket{\varphi_{njm_j}}$, and its magnitude can be estimated as
\begin{eqnarray}
\bra{n'j'}|T^{(k=2)}_e|\ket{nj}
&\sim&\bra{\varphi_{n'j'0}}
(\bm{\sigma}\cdot\bm{p})(\bm{\sigma}\cdot\bm{p})
\ket{\varphi_{nj0}} \nonumber\\
&=& \bra{\varphi_{n'j'0}}
\bm{p}\cdot\bm{p}
\ket{\varphi_{nj0}}
\sim O\left(2m_e(E_{n,j}-m_e)\right) \nonumber\\
&\sim& O\left(m_e^2(Z\alpha)^2\right).
\end{eqnarray}
Consequently, the average transition rate is in the magnitude
\begin{equation}\label{R e estimate}
R^{(e,\pm)}_{i\rightarrow f} \sim m_e^2(Z\alpha)^4 \big|h^0_{R/L}\big|^2\delta(E_f-E_i\mp\omega)
\end{equation}
up to the Clebsch--Gordan coefficients and a dimensionless factor of $O(1)$ depending on $n',j',n,j$.

Because the section rule and Fermi's golden rule for $R^{(e)}_{i\rightarrow f}$ are exactly the same as those for $R^{(\phi)}_{i\rightarrow f}$, the atomic transition induced by $H^{(1)}_e$ again can be understood in terms of gravitons carried by the gravitational wave as discussed in the previous subsection.
As previously, the average number flux of gravitons is proportional to the amplitude square of the gravitational wave, as $R^{(\phi,\pm)}_{i\rightarrow f} \propto \big|h^0_{R/L}\big|^2$ in \eqref{R e estimate}.

From \eqref{R phi estimate} and \eqref{R e estimate}, we have $R^{(\phi,\pm)}_{i\rightarrow f}\propto(Z\alpha)^3$ and $R^{(e,\pm)}_{i\rightarrow f}\propto(Z\alpha)^4$. This tells that, for $Z\alpha\ll 1$, the effect induced by $H^{(1)}_e$ is much smaller than that induced by $H^{(1)}_\phi$ by a factor of $Z\alpha\approx Z/137$.\footnote{For $Z\alpha\sim1$, these two effects can be comparable, but the Schr\"{o}dinger--Pauli approximation is no longer valid.}

\section{Summary and discussion}\label{sec:summary}
By casting the linearized theory of gravity into the tetrad formalism, we are able to formulate the covariant Dirac equation for the electron in a hydrogen-like atom subject to a gravitational wave. The resulting Hamiltonian is given by \eqref{H}, which contains three terms: the original Hamiltonian $H_0$, the Coulomb potential correction term $H^{(1)}_\phi$, and the tetrad correction term $H^{(1)}_e$.

Applying the first-order perturbation theory upon the basis of the energy eigenstates of $H_0$, we can calculate the atomic transition rates induced by $H^{(1)}_\phi$ and $H^{(1)}_e$. Both transitions rates $R^{(\phi)}_{i\rightarrow f}$ and $R^{(e)}_{i\rightarrow f}$ yield the same Fermi's golden rule and selection rule, which can be concisely understood in terms of the notion of gravitons as massless spin-2 particles.

It is remarkable that the transition rates strongly suggest the existence of gravitons, even though the gravitational wave is treated purely as a classical external field. This result is analogous to the familiar case of a hydrogen-like atom subject to an external electromagnetic wave, which yields the selection rule with Fermi's golden rule that suggests the existence of photons as massless spin-1 particles, even if the electromagnetic wave is treated completely classically.
If one takes the concept of gravitons seriously and manages to formulate the interaction between a hydrogen-like atom and the second-quantized gravitational field, it should predict the phenomena of spontaneous emission and stimulated emission of gravitons akin to those of photons except for a different selection rule. Our analysis serves as a stepping stone toward advancing research in this direction.
On the other hand, even if gravitons do not exist, our results remain valid since our analysis is based on very fundamental assumptions without appealing to any quantization of gravitational fields.

Our study of the atomic transitions induced by a gravitational wave is primarily rooted in theoretical considerations. In experimental considerations, measuring these effects poses formidable challenges, not only because the factors $(Z\alpha)^3$ in \eqref{R phi estimate} and $(Z\alpha)^4$ in \eqref{R e estimate} are small, but more significantly, because the amplitude $h_{R/L}^0$ is extremely minute when a gravitational wave reaches Earth.
In outer space, however, the amplitude of a gravitational wave can be substantially strong, potentially leaving discernible imprints on the interstellar medium. Given that the atomic transitions induced by gravitational waves adhere to the specific selection rule distinct from that of electromagnetic radiation, it is possible to detect these effects through a meticulous analysis of the emission or absorption spectra of hydrogen-like atoms in the interstellar medium. Considering that the energy gaps of a hydrogen-like atom are typically in the order of $10^{13}$--$10^{15}$ Hz, this approach offers a distinctive means of probing Ultra-High-Frequency Gravitational Waves (UHF-GWs) in the 10\,THz to $10^3$\,THz range. UHF-GWs are believed to involve new physics beyond the Standard Model, most likely linked to exotic astrophysical entities or early universe events \cite{aggarwal2021challenges}.
In addition to a few detector concepts as surveyed in \cite{aggarwal2021challenges} and the synergy of axion searches and atomic precision measurements as proposed in \cite{domcke2023electromagnetic}, our research suggests a new opportunity for detecting UHF-GWs.

\begin{acknowledgments}
This work was supported in part by the National Science and Technology Council, Taiwan under the Grants MOST 111-2112-M-110-013 and NSTC 112-2119-M-002-017.
\end{acknowledgments}



\input{Atomic_transitions_by_GW.bbl}
%

\end{document}

%% file: Atomic_transitions_by_GW.bbl
%